# Brief wide-field photostimuli evoke and modulate oscillatory reverberating activity in cortical networks

R Pulizzi[1], G Musumeci[1], C Van Den Haute[2-3], S Van De Vijver[1], V Baekelandt[2], M Giugliano[1, 4-5]

[1] *Theoretical Neurobiology & Neuroengineering, University of Antwerp, Antwerp, Belgium;*
[2] *Laboratory of Neurobiology and Gene Therapy, Katholieke Universiteit Leuven, Leuven, Belgium;*
[3] *Leuven Viral Vector Core, Katholieke Universiteit Leuven, Leuven, Belgium*
[4] *Department of Computer Science, University of Sheffield, S1 4DP Sheffield, UK*
[5] *Laboratory of Neural Microcircuitry, Brain Mind Institute, EPFL, CH-1015 Lausanne, Switzerland*

Author's contribution: conceived and designed the experiments: RP, GM, and MG. Performed the experiments: RP and GM. Analysed the data: SVDV and RP. Performed computer simulations: MG. Contributed reagents/materials/analysis tools: CVDH, VB, and MG. Wrote the paper: RP and MG. The authors declare no competing financial interests.

We are grateful to Mr. D. Van Dyck and M. Wijnants for excellent technical assistance, to Drs. J. Couto, D. Linaro, R. Schneggenburger, U. Egert, and S. Marom, for helpful discussions, to Drs. M. Mattia, E. Vasilaki, M. Deger, and two anonymous Reviewers for comments on an earlier version of the manuscript, and to Dr. P. Hegemann for the gift of the CHR2 L132C/T159C-mCherry. Financial support from the 7th Framework Programme of the European Commission (FP7-PEOPLE-ITN "NAMASEN" grant n. 264872, FP7-PEOPLE-IAPP "NEUROACT" grant n. 286403, FP7-ICT-FET project "ENLIGHTENMENT", grant n. 284801), the Interuniversity Attraction Poles Program (IUAP) of the Belgian Science Policy Office, and the University of Antwerp is kindly acknowledged.

Correspondence should be addressed to Dr. Michele Giugliano, Theoretical Neurobiology and Neuroengineering, University of Antwerp, Campus Drie Eiken, Universiteitsplein 1, 2610 Wilrijk, Belgium. E-mail: michele.giugliano@uantwerpen.be

**Abstract**

Cell assemblies manipulation by optogenetics is pivotal to advance neuroscience and neuroengineering. In *in vivo* applications, photostimulation often broadly addresses a population of cells simultaneously, leading to feed-forward and to reverberating responses in recurrent microcircuits. The former arise from direct activation of targets downstream, and are straightforward to interpret. The latter are consequence of feedback connectivity and may reflect a variety of time-scales and complex dynamical properties.

We investigated wide-field photostimulation in cortical networks *in vitro*, employing substrate-integrated microelectrode arrays and long-term cultured neuronal networks. We characterized the effect of brief light pulses, while restricting the expression of channelrhodopsin to principal neurons. We evoked robust reverberating responses, oscillating in the physiological gamma frequency range, and found that such a frequency could be reliably manipulated varying the light pulse duration, not its intensity. By pharmacology, mathematical modelling, and intracellular recordings, we conclude that gamma oscillations likely emerge as *in vivo* from the excitatory-inhibitory interplay and that, unexpectedly, the light stimuli transiently facilitate excitatory synaptic transmission. Of relevance for *in vitro* models of (dys)functional cortical microcircuitry and *in vivo* manipulations of cell assemblies, we give for the first time evidence of network-level consequences of the alteration of synaptic physiology by optogenetics.

**Keywords**

Optogenetics; substrate-integrated microelectrode arrays; *in vitro* cortical networks; gamma rhythms; oscillations; reverberating activity;

**Introduction**

Since first expressing channelrhodopsin in neurons [1], optogenetics [2] raised immense interest as one of the most promising techniques in neuroscience [3-8]. Optogenetics enables the manipulation of the activity of genetically-identified neurons with unprecedented temporal resolution [9-14]. Therefore, both in fundamental research and neuroprosthetics, the idea of regulating neuronal firing by light (e.g. in a closed loop) quickly emerged as a pivotal causative approach to advance our correlative understanding of brain (dys)functions. Nevertheless, the limits and most useful features of the stimulation (e.g. intensity, time course, frequency, temporal pattern, etc.), to be exploited in a control system, have not been exhaustively explored, but see [15,16].

Despite the low spatial resolution in most *in vivo* applications, but see [17,18], as light broadly addresses the feed-forward projections expressing an opsin, the target neuronal populations can be efficiently controlled [5,19,20]. In this case, optogenetic control of downstream neuronal firing is straightforward and irreplaceably elegant and simple. However, when recurrent collaterals expressing an opsin dominate, e.g. as in the cortex, it is expected that light stimuli elicit additional reverberating responses, due to local recurrent pathways. These responses arise as second-order effects from the recruitment or suppression of neuronal activity and may reflect unanticipated time-scales, and complex dynamical microcircuit properties. These add complexity to the interpretation of optogenetic stimulation and to its use for closed loop control, particularly *in vivo* where they might be hard to isolate and study in depth. A complementary approach for dissecting reverberating activity may come from using of reduced *in vitro* experimental preparations [21-23], offering substantial advantages over *in silico* simulations. In fact, while addressing the impact of recurrent connectivity greatly benefits from detailed and simplified neuronal models [24,25], introducing opsins's biophysics into large-scale network simulations has been infrequent so far, but see [26,27,28]. In addition, most of Hodgkin-Huxley-like models fail to capture the extended dynamical properties of neurons and synapses [29,30].

Here we explore *in vitro* the collective response evoked by light in recurrent microcircuits, employing mature cultured neurons, dissociated from the rat neocortex, and let developing *ex vivo* on microelectrode arrays (MEAs) for several weeks. As neurons reorganized into functional recurrent networks [31,32], MEAs allowed non-invasive and chronic access to their coordinated activity [33-36]. We selectively targeted the expression of a variant of channelrhodopsin in principal neurons and delivered brief wide-field light stimuli, while recording spiking activity from the MEA microelectrodes.

We observed that evoked network responses outlasted the stimuli, as for focal extracellular electrical stimulation [37,38], and also contained transient oscillations in the mean firing rate with frequency in the gamma range [75]. Unexpectedly, the duration of the light pulses reproducibly altered the temporal course of the reverberating response and modulated the frequency of its oscillations from ~50 to ~150 cycles/s. A simple mathematical model suggested a possible synaptic mechanism underlying our observations, which was later confirmed by performing intracellular recordings and monitoring synaptic release events immediately before and after the light stimuli. All in all, these results enhance our understanding on how reverberating activity can be manipulated, offers an experimental model to study gamma-range rhythms in a dish, and provide for the first time, evidence of unexpected network dynamical consequences of the alteration of synaptic physiology by optogenetics.

**Results**

We cultured large-scale networks of rat primary cortical neurons, over 57 arrays of microelectrodes (MEAs) integrated in transparent glass substrates. Over four weeks *in vitro*, neurons developed functional synaptic connections and reached a mature pattern of spontaneous activity [34]. By detecting action potentials recorded at each microelectrode, we investigated the collective neuronal responses upon brief wide-field light stimulation (Fig. 1A). In 39 MEAs, we transduced cells by AAV to express a variant of channelrhodopsin (CHR2 LC-TC) fused to a red-fluorescent protein (mCherry), targeting selectively CaMKIIα-positive cells (i.e. putative glutamatergic neurons) [53] (Fig. 1D). This mutant opsin responds to blue light with stronger currents, reduced proton permeation, and enhanced calcium ($Ca^{2+}$) selectivity, compared to its wild type [40]. Immunocytochemistry confirmed our desired high level of opsin expression, transduction efficiency (i.e. ~80% of all neurons), and low apoptosis (not shown). Routine MEA recordings showed no differences with control cultures in terms of network development and cell survival, inferred by analyzing spontaneous bursting activity (Fig. 1B-C) [22,43] and by quantifying the number of active microelectrodes (i.e. those detecting >1 spike in 50s; Fig. 1C).

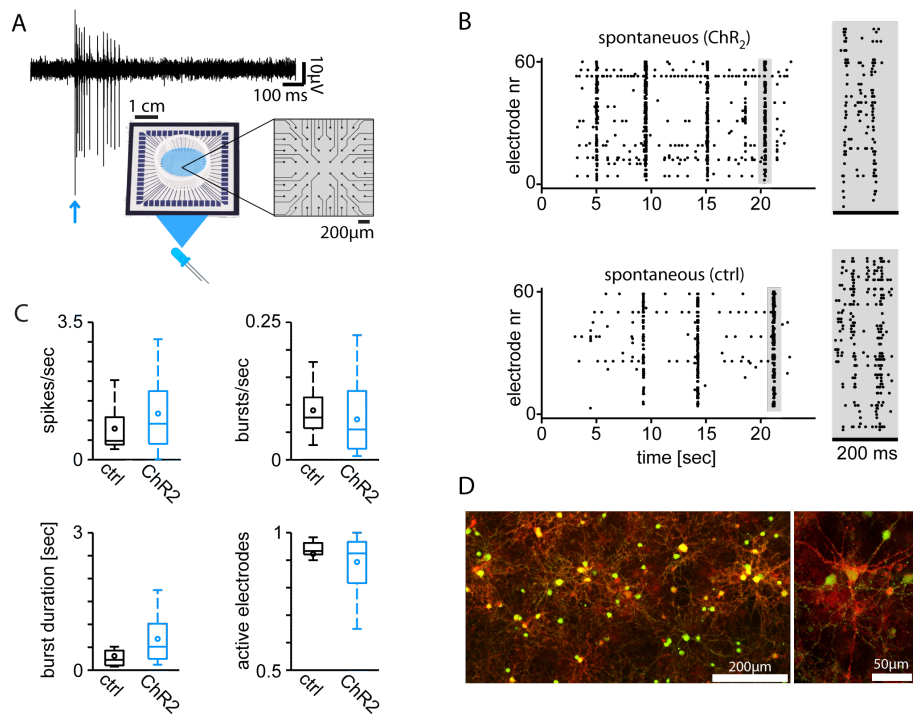

**Figure 1: Opsin expression in cultured neurons does not alter the spontaneous network electrical activity.** Transparent glass microelectrode arrays (MEAs), used as substrates for long-term primary cortical cultures, were employed to record light-evoked (A) and spontaneous (B) neuronal electrical activity, 28 days after plating. Brief light pulses were delivered wide field by a computer-controlled LED, focused on the entire inner area of the MEA (A), and found to elicit neuronal spiking activity analyzed from extracellular raw voltage signals. Transduction of ChR2 and mCherry by AAV vectors, with a CaMKIIα promotor, altered neither the spontaneous network bursting (B), nor induced significant changes in neuronal firing rate, burst rate and duration, or in the number of active microelectrodes (C, quantified over 11 control and 46 transduced MEAs). NeuN immunostaining (green) and fluorescent mCherry (red) imaging highlighted (D) the ChR2 expression in putative excitatory neurons, with spatially unrestricted presence over the cell surface.

**Light-evoked network responses.** We employed a computer-controlled blue LED to deliver brief wide-field pulses of light over the inner bottom area of each MEA. Simultaneously we recorded the neuronal spike responses, detected extracellularly at each of the 59 microelectrodes of the MEA (Fig. 1A, inset). The duration of light pulses was systematically varied in the set {0.1, 0.5, 1, 2, 5, 10, 20} ms, while the light intensity was kept constant to its full power value (i.e. 2 mW/mm$^2$ at the sample). This allowed us to vary reliably the amplitude of light-activated CHR2 currents, as validated by patch-clamp experiments (see Supplemental Figure S1). Each pulse was repeated 60 times at a very low rate (0.2Hz), thus reducing the likelihood of short- and long-term plastic changes in neuronal responses [55].

Regardless of the pulse duration, the evoked network response outlasted the stimulus and exhibited an early and a late component, as apparent in the shape of the peri-stimulus histogram (PSTH) of the spike times (Fig. 2A). Control cultures displayed as expected no evoked response (not shown). Similar to electrically evoked responses [37], the two components were related to the direct neuronal activation and the reverberating interactions mediated by recurrent connectivity. When the intensity and not the duration of the light pulse were changed, considerably less stable and reproducible late responses were observed (not shown), correlating with weaker early responses.

The direct component, occurring in the first few milliseconds after the stimulus onset, originated from the intense sudden depolarization caused by the CHR2 activation. It had low temporal jitter across microelectrodes and repetitions, and was followed by a transient decrease of activity. The reverberating response arose several tens of milliseconds after (i.e. 30-50ms), as a renewed progressive build up and then exhaustion of neuronal firing. It lasted up to several hundreds of milliseconds and reflected in part the slow deactivation kinetics of CHR2 LC-TC, and in part the effect of recurrent network interactions. As a consequence of the high reliability of the first evoked 1-2 spikes/electrode, observed intracellularly in individual neurons (not shown), the early peak in the PSTH was rather sharp and highly reproducible, with its amplitude and latency not significantly affected by the total pharmacological blockade of synaptic transmission. Amplitude and latency of the early peak in the PSTH did not significantly correlate with the light pulse duration (Fig. 2B). Instead, the peak amplitude and latency of the reverberating response were affected by the total pharmacological blockade of synaptic transmission (not shown) and significantly correlated with the light pulse duration (Fig. 2C-D; $r = 0.42$ and $r = -0.36$, respectively, $p < 0.0001$): the longer the pulse, the weaker and more delayed the reverberating response. Differences in peak amplitudes and latencies, upon varying pulse duration, were highly significant ($p<0.01$) across the entire dataset ($n = 39$ MEAs), but only for the reverberating responses, not for the early responses.

Examining the evoked responses in more details, employing higher temporal resolution (i.e. 1ms bins), we also found signatures of global oscillatory activity in the PSTHs, with striking similarity to physiological gamma-range oscillations. Figure 3A illustrates, for a representative MEA, the spikes evoked by the stimulus as a raster plot of their times of occurrence and its corresponding PSTH, across 59 microelectrodes and 60 repetitions. The estimate of the time-varying power spectrum of the PSTH (Fig. 3B) clearly revealed the appearance of a fading oscillation with frequencies in the gamma range (i.e. ~[40; 100] cycles/sec), as conventionally measured 75ms after the stimulus onset, and qualified as *dominant* in terms of signal-to-noise ratio (see the Methods). Across all MEAs tested, oscillations completely faded away ~200ms after the onset of the stimulus, and generally slowed down at a rate of ~0.3cycles/sec per msec over time. Most importantly, the frequency of the oscillations significantly correlated with the pulse duration ($r = 0.55$, $p<0.0001$): the longer the pulse, the faster the *dominant* oscillation (Fig. 3C-D).

In addition, an initially low variability of the spike count, estimated by the Fano factor at each microelectrode across repetitions [56], immediately raised to higher values (i.e., from 0.4 to 1) within 20-50ms from the stimulus onset (not shown). Moreover, the occurrence time of the spikes recorded at individual microelectrodes showed no entrainment with the oscillation cycles of the PSTH over time, but rather irregular discharges with spike rates generally below 50 spikes/sec (not shown). Overall, these results suggest that excitatory-inhibitory feedback, rather than a beating-wave arising from spike synchronization drift over time, is the prevailing mechanism of the observed oscillations.

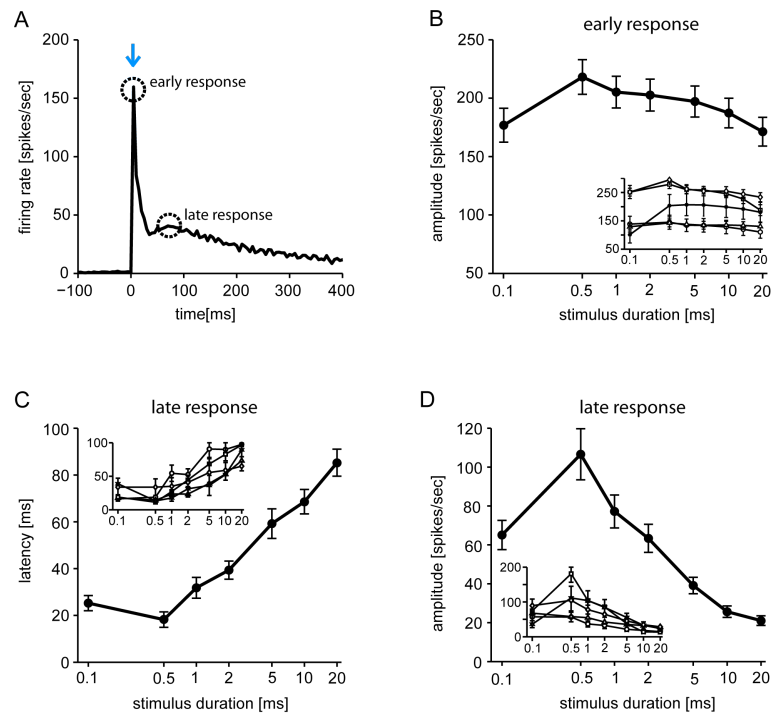

**Figure 2: Light-evoked network responses.** Brief light pulses evoked network-wide spiking responses, only in cultures transduced by AAV. Responses were quantified over 60 repetitions by computing the peri-stimulus histogram of spike times (PSTH) (A), detected over 5 ms bins across all microelectrodes (Fig. 1A). In all experiments, the time course of the PSTH was stereotyped and consisted of an early and late phase, similarly to what was described for extracellular electrical stimulation. While intensity was fixed, the duration of light pulses was changed systematically and it was unexpectedly found to significantly modulate the late (C-D) but not the early phase of the PSTH (B). Longer pulses significantly ($p<0.001$) delayed (C) and weakened (D) the late response peak, without affecting the early phase (circles in A). Evaluating the non-parametric Kendall's rank coefficient, a significant correlation ($p<0.0001$) was only found between pulse duration and late-peak latency (0.47) or between pulse duration and late-peak amplitude (-0.36). Panels B-D summarize the results over 39 MEAs, while the insets show averages across several MEAs, over five distinct sets of sister cultures, remarking the consistency of the findings across different biological preparations.

**Similarity of evoked and spontaneous network events.** Inspired by a previous report [68], we analysed in more detail the time course of the population mean firing rate during spontaneous network-wide events (i.e. network bursts - Fig. 1B) [34,43], in the absence of any light stimulation. While for evoked responses the spike times were related to the stimulus onset and their histograms averaged accordingly across repetitions (Fig. 2A), for spontaneous events no temporal reference frame exists by definition. We thus rigidly shifted all spikes, detected simultaneously across the microelectrodes during each spontaneous burst, by a common interval adapted to maximise the firing rate time course similarity across bursts [68] (see the Methods). Unexpected temporal features then emerged in the average firing rate, which could not be revealed by routine analysis methods based on firing rate peaks [43] or threshold-crossing alignment [43] of spontaneous bursts. We found that some but not all MEAs showed prominent signatures of spontaneous transient oscillations in the time course of the firing rate within a network burst, occurring in the same physiological gamma frequency of evoked rhythms (Fig. 4), similarly to the light evoked responses.

**GABA$_A$ receptors are necessary for the evoked global gamma oscillations.** In order to investigate the synaptic mechanisms of the evoked oscillations, we pharmacologically abolished fast inhibitory synaptic transmission. As inhibition is necessary for gamma-range rhythmogenesis *in vivo* [75], we tested the hypothesis that excitation-inhibition interplay sustains light-evoked oscillations *in vitro*. When a competitive selective blocker of GABA$_A$ postsynaptic receptors was bath-applied, not only the spontaneous network activity significantly increased in terms of firing and burst rate (Fig. 5A) as described [34,57,58]. In addition, while light-evoked responses still consisted of an early and a late component, no oscillatory activity occurred as no single peak in the power spectrum of the PSTHs qualified as a *dominant* frequency (n = 17 MEAs) (Fig. 5B-C). Hence, the pharmacological manipulation of network activity corroborates the interplay between excitatory and inhibitory neurons, necessary for the light-evoked network oscillations.

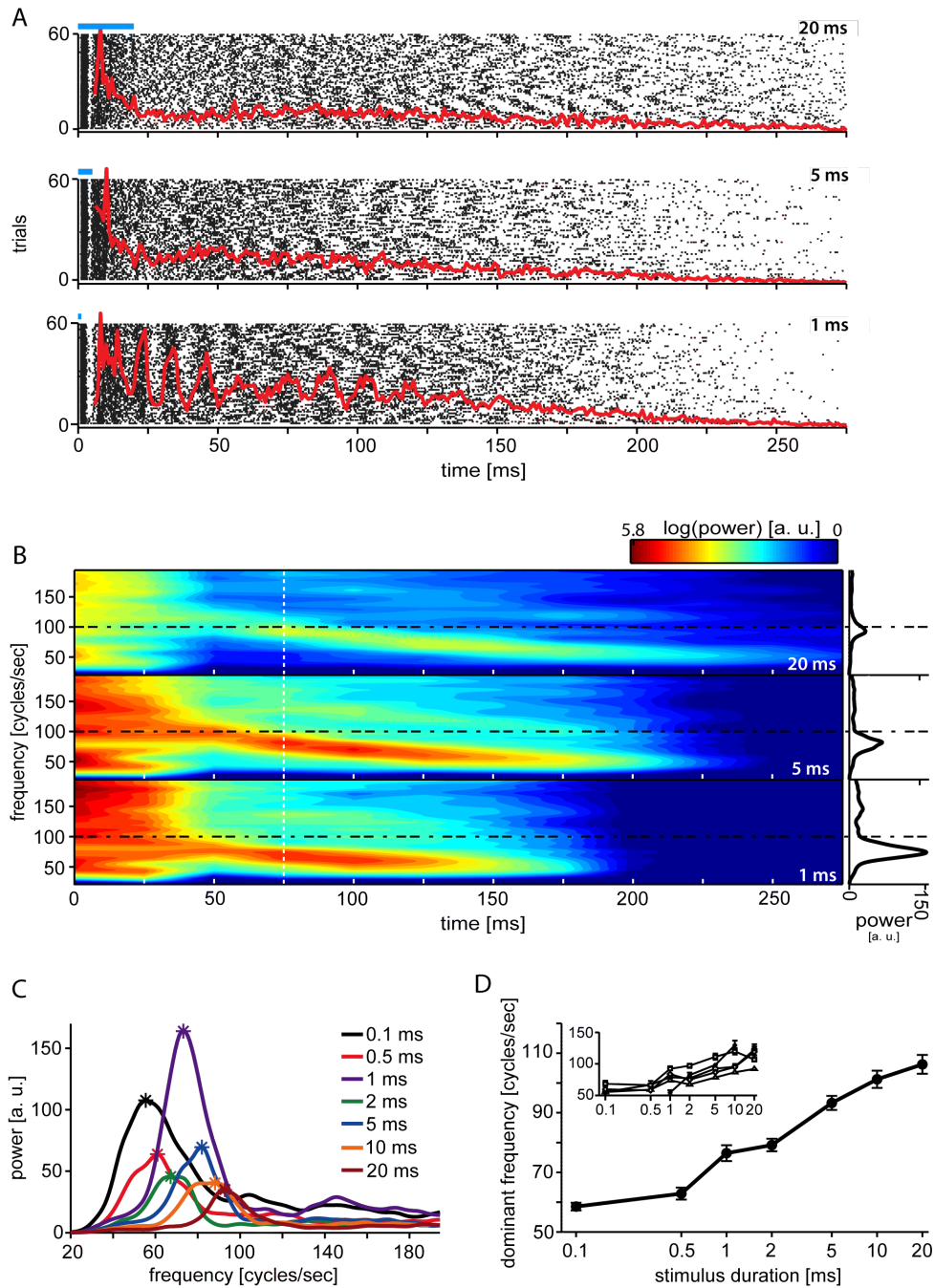

**Figure 3: Oscillatory light-evoked responses are modulated by the duration of light pulses.** Typical light-evoked responses are plotted for three light pulse durations (i.e., 1, 5, and 20 ms) as (A) spike-time rasters (black dots) over all electrodes and trials, and as spike-time PSTH (red traces), computed over 1 ms bins and peak normalized. The presence of transient oscillations was quantified by estimating the time-varying power spectrum of each PSTH (B, left panels) and further slicing it 75 ms after the stimulus onset (B, right panels, white dashed line; see the Methods). A clear dominant frequency (see Material and Methods) was reliably observed in 37 out of 39 MEAs (for 1, 2, and 5 ms long stimuli) or in 30 out of 39 MEAs (for 0.1, 0.5, 10, and 20 ms long stimuli). The dominant frequency showed a progressive drift towards slower oscillations over time (B), and it significantly varied for increasing values of the pulse durations

(p<0.001; C-D). Evaluating the non-parametric Kendall's rank coefficient, a significant (p<0.0001) correlation was found between dominant frequency and pulse duration (0.55).

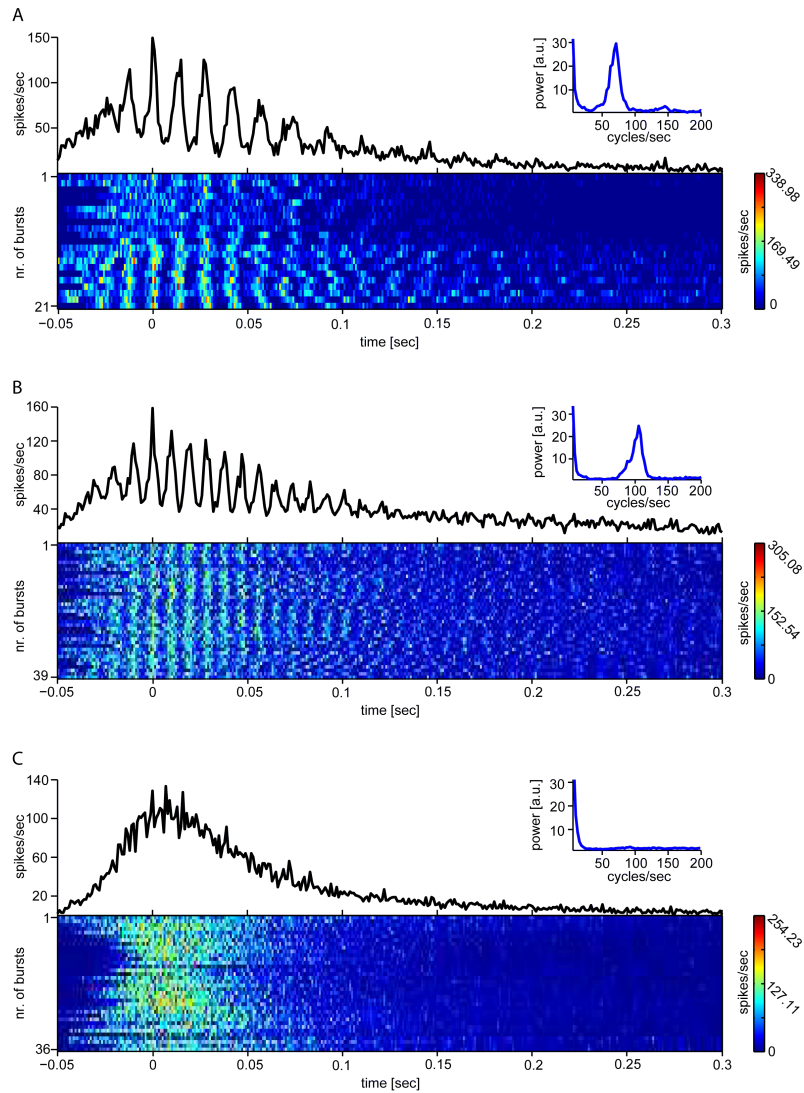

**Figure 4: Oscillatory firing activity during spontaneous network bursting.** By aligning for maximal similarity the spike trains, recorded by MEA microelectrodes for each spontaneous network burst, and averaging as in [68], some (A-B) but not all MEAs (C) showed prominent signatures of oscillations in the same physiological gamma frequency of evoked rhythms. The power spectra (insets) further quantify that, when occurring, oscillations were dominated by physiological gamma range frequency.

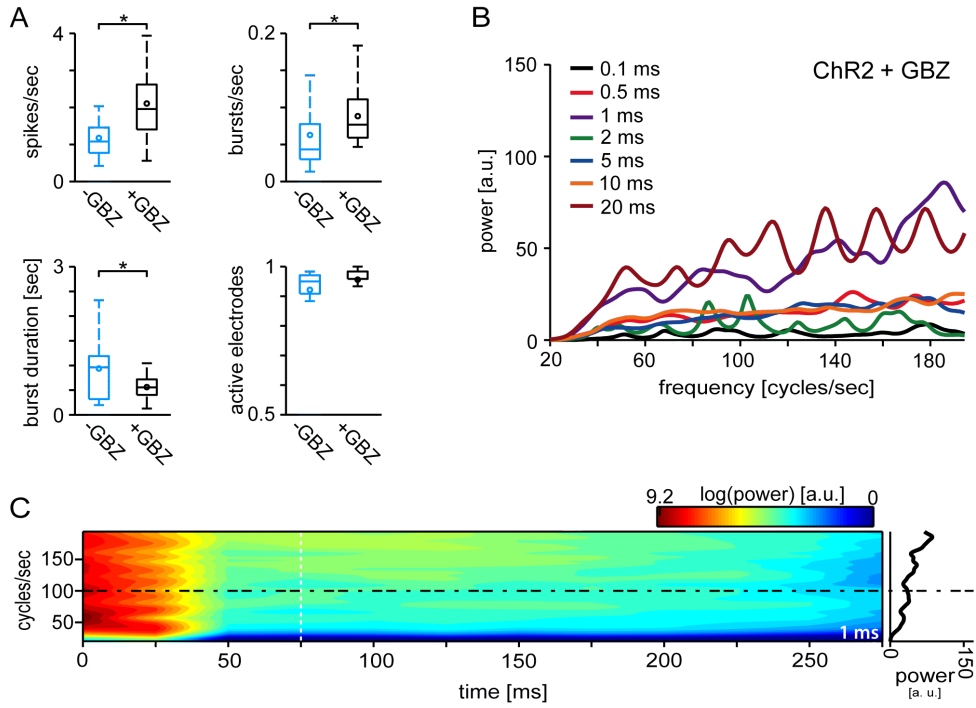

**Figure 5: Inhibition is necessary for light-evoked oscillations.** The pharmacological blockade of GABA$_A$ receptors disinhibited the network and altered quantitatively but not qualitatively the spontaneous activity, significantly (p<0.05) increasing the burst rate and decreasing the burst duration (A). However, light-evoked responses were qualitatively affected by network disinhibition as they did not contain oscillatory components (n = 17 MEAs), as revealed by the power spectrum analysis (see Fig. 3) and the lack of a dominant oscillation frequency (B-C).

**Oscillations *in silico* by the excitatory-inhibitory feedback loop.** In order to further investigate the consequences of the excitatory-inhibitory feedback loop, we considered a classic firing-rate mathematical model [50]. The *in vitro* cortical networks considered here are composed by both excitatory and inhibitory neurons, showing a prominent pattern of recurrent synaptic connectivity [34]. We therefore defined a model composed of an excitatory population and an inhibitory population with recurrent excitation and feedback inhibition (Fig. 6A), and we analysed and numerically simulated its response to an external constant input. As excitatory neurons express ChR2 in the experiments, only the excitatory population was activated by the external input. After linearization of eqs. 4-6, the approximate conditions for stable oscillatory activity could be formulated in terms of average synaptic efficacies. Assuming identical strength of excitatory synapses for simplicity, i.e. $J_0 = J_{ee} = J_{ie}$, recurrent excitation should not compromise the stability of the system, i.e. $J_0 < 2\alpha_e^{-1}$, but must be strong enough to recruit the inhibitory population, i.e. $J_0 > (\alpha_e \alpha_i J_{ei})^{-1}$. This range of values for $J_0$ exists if the feedback inhibition is sufficiently strong, i.e. $J_{ei} > \alpha_i^{-1}$ (Fig. 6C). Under these hypotheses, oscillations occur at a frequency *f*, given as the imaginary part of the conjugate complex eigenvalues of the connectivity matrix A (eq. 8), irrespective of the amplitude and duration of the external constant input [47]:

$$f = (2\pi)^{-1}(2\tau)^{-2}\sqrt{4\alpha_e \alpha_i J_{ei} J_0 - (\alpha_e J_0)^2} \qquad (1)$$

Equation 6 implies that feedback inhibition is necessary in our simplified *in silico* scenario (Fig. 6C), as for $J_{ei} = 0$ the term below the square root would be negative, and supports the interpretation of the experimental results obtained under blockade of GABA$_A$.

Importantly, any (unmodelled) mechanism that should increase the efficacy of excitatory synapses $J_0$, e.g. as in short-term synaptic facilitation, would be sufficient to increase *f* and thus speed up the global oscillations (Fig. 6B). Similarly, any (unmodelled) mechanism that should decrease $J_0$, e.g. as a recovery to its resting value following an intense facilitation, would account for the progressive slowdown of the oscillations through time.

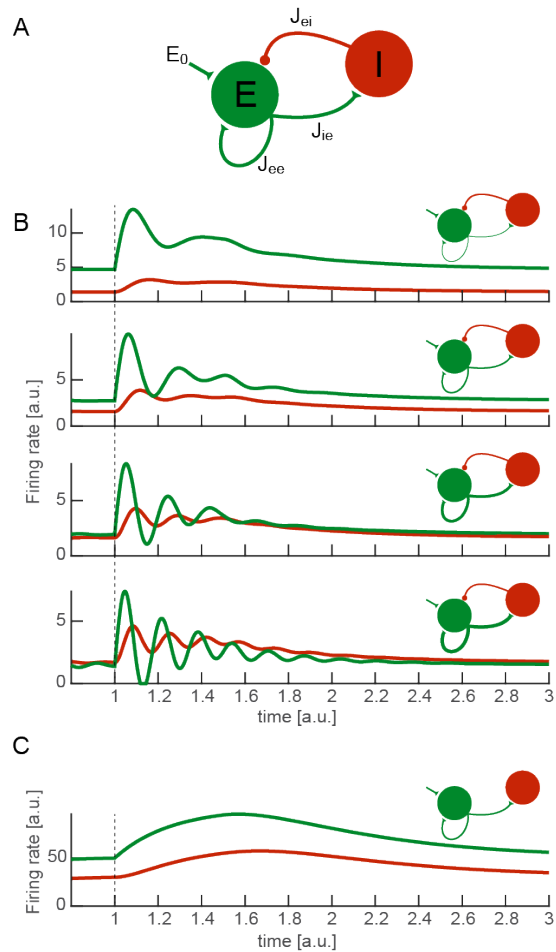

**Figure 6: A firing rate model links evoked oscillations to excitatory-inhibitory interactions.** By a firing-rate mathematical model (A) we qualitatively reproduced oscillatory transient activity *in silico* (B). Recruited by the external stimulation of excitatory neurons, the negative feedback of the inhibitory loop generated a well-known alternating pull able to decrease neuronal firing. The red and green traces represent qualitatively the mean firing rate of inhibitory and excitatory neurons, respectively. Altering the efficacy of synaptic connections in a series of numerical simulations (B), resulted in different frequencies of the emerging network rhythm: increasing the strengths of excitatory connections (i.e. $J_{ie}$ and $J_{ee}$) increased the frequency of the oscillations. As in the experiments (Fig. 5), inhibition was necessary for the oscillations, as they completely vanished upon removal of the inhibitory feedback (C, $J_{ei} = 0$). However, the model could account per se neither for the dependency of the oscillation frequency on the duration of the light pulse, nor for its drift over time. As a small increase in the average excitatory synaptic efficacy is sufficient *in silico* to speed up oscillations, the duration of the light pulse might indirectly affect synaptic efficacy in the experiments, as in short-term facilitation.

**Short-term facilitation of synaptic release probability at excitatory synapses.** The insights obtained with the model suggest that light, and its duration, might proportionally affect the excitatory synaptic efficacy. We therefore hypothesised that the link between light duration and oscillation frequency might involve the influx and transient accumulation of $Ca^{2+}$ in the excitatory synaptic boutons, as in the residual $Ca^{2+}$ hypothesis for short-term synaptic facilitation [59,60]. This hypothetical link, occurring by the activation of CHR2 LC-TC and voltage-gated calcium-channels, would lead to a (transient) increase of the action potential-dependent excitatory synaptic efficacy (i.e. analogous to $J_0$ in eq. 1) and it would compete with the endogenous $Ca^{2+}$ recovery processes. That the synaptic release probability should have been reversibly affected by the light stimulus, and depend linearly on its duration was a prediction of our model.

Intending to test our prediction, we performed whole-cell intracellular recordings from the soma of neurons not expressing CHR2 LC-TC (n = 5 cells), by selecting mCherry-negative cells through videomicroscopy. Then, we blocked action potential (AP) initiation by bath application of a selective blocker of voltage-gated sodium channels (i.e. tetrodotoxin, TTX). Recording in voltage-clamp, we quantified number and instantaneous rate of AP-independent excitatory postsynaptic events [61,62]. Such events are the result of neurotransmitter release, occurring in connected presynaptic boutons spontaneously or reflecting a transient increase in the free $Ca^{2+}$ concentration in the boutons [63,64]. We found that immediately after the light stimulus, the presynaptic release probability abruptly and reversibly increased (Fig. 7C), and it was linearly correlated at its peak with the duration of the light pulse (Pearson's r = 0.7, p<0.0001). This observation is compatible with the residual $Ca^{2+}$ accumulation in the synaptic terminal. In addition, the presynaptic release probability reversed significantly over time, exponentially decaying with a time constant of ~200ms (Fig. 7D), similarly to the recovery from short-term facilitation.

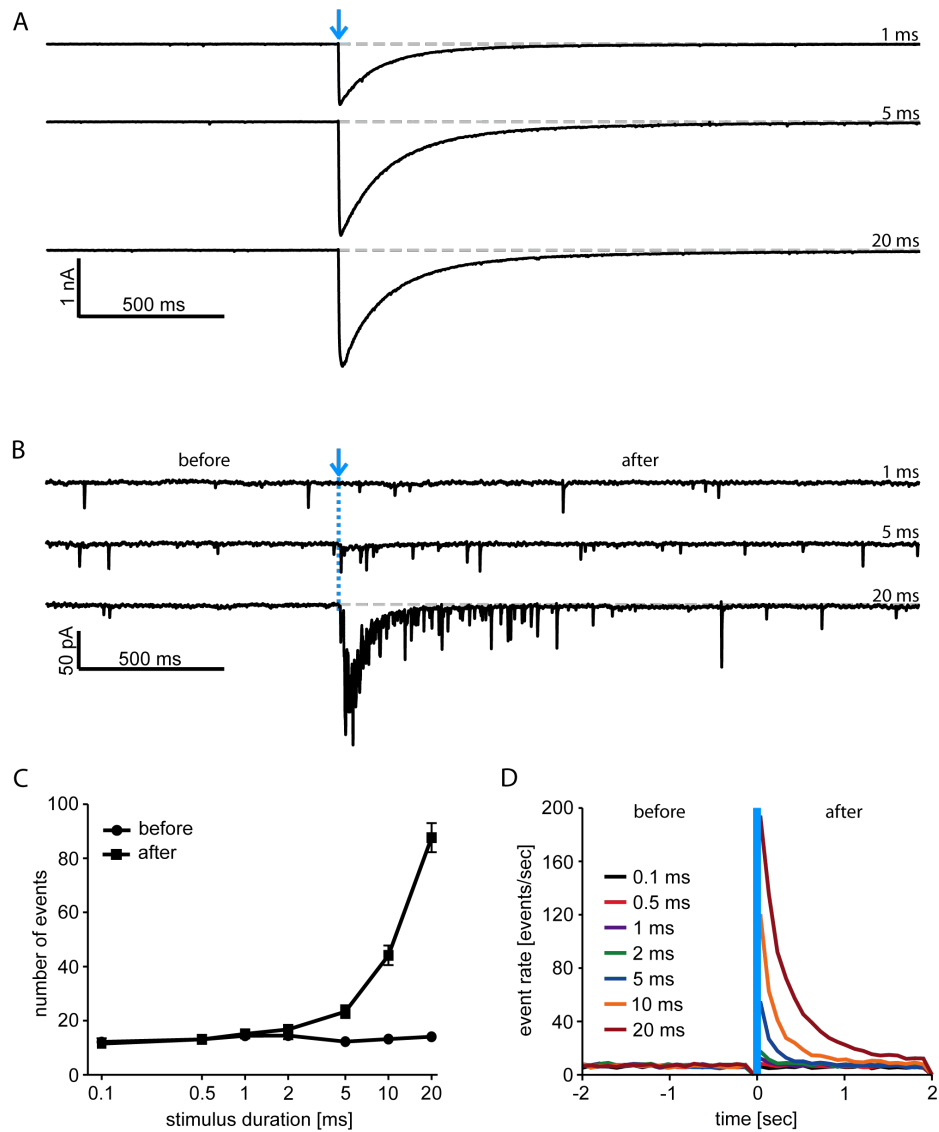

**Figure 7: Brief light pulses transiently increase the rate of AP-independent synaptic release events.** Intracellular whole-cell voltage-clamp recordings were performed while blocking sodium-currents by TTX, suppressing initiation and propagation of action potentials (APs). The time course of intrinsic and synaptic currents, elicited by brief light pulses (blue arrow and dotted line), reveal fast on-kinetics and slow off-kinetics of ChR2-LCTC (A), which strongly prevails on synaptic currents in mCherry-positive neurons. Recording from mCherry-negative neurons, only synaptic currents are observed (B). The light pulses induced a transient increase in the rate of spontaneous synaptic release event significantly correlated with the duration of the pulse (0.7, p<0.0001) (C). The time course of the release events, quantified in 100 ms bins (D), decayed exponentially with a time constant of ~200 ms, roughly independent on the pulse duration.

**Discussion**

Our results show that dissociated neuronal cultures are an interesting model for the study of gamma-range oscillations and are useful to investigate some non-trivial consequences of wide-field stimulation in recurrently connected neurons. Through this experimental model, we report for the first time how to robustly evoke *in vitro* an oscillatory reverberating network response, and reproducibly modulate its frequency, within a range of physiological importance. Many earlier studies reported on spontaneous and evoked oscillations *in vitro* [65-68], but none described a fine experimental control of its features. This might prove very relevant for *in vitro* models of (dys)functional cortical microcircuitry [69,70] and for *in vivo* optogenetic manipulations of cortical cell assemblies.

We interpreted the emergence of gamma-range damped oscillations from the known interplay between excitation and inhibition, supported by pharmacological experiments and mathematical modeling. An alternative interpretation is offered from computational studies: the firing rate of a population of weakly-coupled neurons show damped oscillations when all cells respond simultaneously and regularly to the same strong stimulus onset [71-73]. Roughly the opposite of a beating-wave, the damped oscillations observed in our experiments might be the consequence of jitter and progressive random shifts of otherwise initially synchronous and regular spike trains. The onset of the light pulse might transiently reset and *de facto* synchronize the firing of the majority of the cells. Neurons would then start to fire regularly and roughly synchronously, while progressively going out of phase due to distinct membrane properties, heterogeneous CHR2 expression, synaptic interactions, etc.

In such a scenario however, single-neuron firing rates would roughly match the global rhythm of PSTH, with an action potential per cycle. Then, the blockade of feedback synaptic inhibition would not necessarily disrupt the oscillations. For these reasons, and for the physiological range of rhythms observed, the theory of sparsely synchronized oscillations proposed in excitatory-inhibitory network architectures to explain the emergence of gamma oscillations *in vivo* [74-76], might better describe our *in vitro* data. In fact, the pharmacological blockade of $GABA_A$ receptors completely abolished the oscillations in the PSTH, the spike count across trials showed high variability, and the temporal pattern of firing detected at single microelectrodes appeared to be not completely entrained with the oscillation cycles.

We note however that the blockade of $GABA_A$ receptors also removes an inhibitory tone in the excitatory neurons, steepening their frequency-current curves and depolarising their resting potential. It is then possible that these two effects might alter the conditions for oscillations emergence, in a scenario where excitation alone generates the rhythm. Future experiments, mimicking the inhibitory tone in disinhibited networks by Optogenetics (e.g. expressing CHR2 and Archaerhodopsin in excitatory neurons), complemented by accurate biophysical spiking neural model, will be needed to explore the validity of this alternative hypothesis.

In summary, the activity we observed *in vitro* captures many of the feature of *in vivo* gamma-range rhythms, such as its short-life, the requirement for an intact excitatory-inhibitory loop, and concurrence with irregular single-neuron firing [75].

To further strengthen this possible conclusion, it would be intriguing in the future to measure the dynamical response [77] of both mCherry-positive/negative cells *in vitro*, as well as of short-term synaptic plasticity [74-76]. Combining these quantitative descriptions together, as described by Wang (2010), one could predict whether or not gamma-range global rhythm can be sustained by recurrent connectivity at the observed firing rates. This might serve as an experimental validation of the theory of sparsely synchronized oscillations [74-76], ultimately linking intrinsic and synaptic properties to emerging oscillations in the same *in vitro* biological preparation.

An important finding of our experiments is the proposed causal relationship between duration of light pulses and excitatory synaptic efficacy. While the probability of synaptic release was already shown to be altered upon optogenetic stimulation [78], here we describe a network-level correlate of this effect for the first time. All our experimental results are consistent with our hypothesis of light-induced short-term facilitation of excitatory synaptic efficacy. This requires the reasonable assumption that the transient and reversible increase in $Ca^{2+}$, observed in terms of an enhanced quantal synaptic release, affects the action potential-dependent synaptic release probability, as in the residual $Ca^{2+}$ hypothesis for explaining activity-dependent short-term facilitation [59,60]. The impact of CHR2 on $Ca^{2+}$ transients has already documented postsynaptically in dendrites and spines [78,79], although linked to voltage-gated calcium-channels activation. Here we advanced the hypothesis that direct calcium influx through CHR2 at presynaptic boutons is responsible for the observed phenomenon.

Ultimately, a full spiking neuron model, including a modified version of the Tsodyks-Markram model of short-term synaptic plasticity [80] would be relevant to test the CHR2 LC-TC contribution on synaptic variables. This model would extend the very basic predictions offered by our analysis of the mean-field model (Fig. 6; eq. 1). Nonetheless, the rate model we examined was simple enough to test the excitation-inhibition feedback loop hypothesis for the oscillations, as well as to explicitly suggest a candidate observable for the additional intracellular experiments of Fig. 7A-B.

An alternative explanation of the relation between the light pulse duration and oscillation frequency might involve the transient alteration of the excitability of (excitatory) neurons expressing CHR2. In fact, as CHR2 LC-TC takes time to deactivate, the total effective membrane ionic conductance in those neurons is larger immediately after the stimulation than before it. This effect would however not account for the increase in the frequency with an increase in the duration of the light pulses. An increase in the effective membrane conductance is known to decrease the slope of single-cell input-output transfer function [81], thus leading to opposite consequences than those observed, for the frequency $f$ of oscillation (see $\alpha_e$ in eq. 1). In addition, while our hypothesis of presynaptic $Ca^{2+}$ accumulation explains why progressively longer pulses result in faster oscillations, a stimulus-dependent change in the total membrane conductance per se would not explain any temporal integration of the stimulus feature. Its value would be dependent only on the time since the onset of CHR2 activation, irrespective of the pulse duration.

We also remark that the observed slowdown of the oscillation instantaneous frequency (Fig. 3B) is consistent with our hypothesis of a transient accumulation of $Ca^{2+}$, and its expected depletion during a subsequent recovery phase, as well as with the short-term synaptic depression. As mCherry was expressed over the entire neuronal morphology (Fig. 1D), the spatially unrestricted expression of CHR2, presumably even at excitatory synaptic boutons, supports the $Ca^{2+}$-accumulation hypothesis. Thus, as the brief light pulse instantaneously primed a synapse, subsequent (endogenous) presynaptic action potentials might find easier to release from the same synapse, as in short-term (spiking) activity-dependent facilitation [59,60].

Our intracellular recordings performed under TTX (Fig. 7B-C) however represent only indirect experimental evidence of the presynaptic facilitation of synaptic efficacy. Intracellular recordings, performed simultaneously from a pair of synaptically connected neurons, would be necessary to quantify whether action potential-dependent synaptic release appears potentiated immediately after delivering light stimulation. While wide-field photostimulation might not be entirely appropriate in that context, a similar experiment might conclusively prove that CHR2 affects recurrent synaptic transmission by altering synaptic efficacy. In addition, with recently developed opsins e.g. Chronos, [82] with altered $Ca^{2+}$ permeability or faster closing kinetics, we predict that the dependency of the oscillation frequency on the light stimulation duration should change dramatically.

We finally note that the validity of our linear analysis of the mathematical model does not extend to very low firing rate regimes, due to the non-linear response properties of neurons near their firing threshold. In such a regime, oscillations take a more complicated form and their frequency is also determined by the balance between excitation and inhibition.

The finding of a signature of gamma rhythms also during spontaneous bursting, in some but not all MEAs (Fig. 4), is also of great significance to interpret the light activated responses. Despite based on very short (i.e. 5 mins) recordings for each MEA, originally meant as a viability control, this finding generalizes to mature large-scale networks what was first reported in developing small neuronal circuits [68]: an "innate" mode of operation of recurrent neuronal circuits. In addition, our observation indirectly confirms a broad heterogeneity in the excitation-inhibition ratio in cortical networks, reflected in the variability of the spontaneous rhythms frequency from ~50 to ~100 cycles/s (Fig. 4) (or the lack of it) [88]. In addition, transitions were sometimes observed (e.g. Fig. 4A, lower panel) between non-oscillating and oscillating modes of intra-burst activity, consistent with the broad range of time-scales for single-neuron and network excitability [88]. Finally, our results further support the dynamical character of such an excitation-inhibition ratio [88], as it can be readily manipulated by internal or external inputs (i.e. light pulses).

Overall, in line with many previous studies [4,8,14,83], our results stress the necessity to take into account the possible impact on recurrent connectivity when designing optogenetic experimental protocols. Moreover, we emphasize the importance of further developments in the field, particularly confining the opsin expression in specific regions of a neuron [84], based on exact experimental needs. In fact, the light-induced short-term facilitation of excitatory synaptic efficacy might prove to be a bug for some applications and not a feature. Ultimately, besides further investigating and clarifying the impact of optogenetic manipulations in large-scale recurrent cortical networks, particularly those affecting synaptic release probability, our work provides a means to manipulate the network responses, for the benefit of future principled design of closed-loop control paradigms.

Large-scale cultured networks are acknowledged as an experimental preparation, where structural, dynamical, and plastic properties of *in vivo* cortical networks may be effectively investigated [22, 32, 34, 38, 43, 89-95]. Due to their inherent 2-dimensional geometry, lower effective cell packing density, and random arrangement of cells' morphology in the plane, they cannot capture faithfully the features of electrical local field potentials (LFPs) as recorded in 3-dimensional layered structures as brain slices and *in vivo* [96]. While our (evoked or spontaneous) *in vitro* gamma oscillations occur prominently in the network-wide probability of firing (Fig. 3A), intact brain rhythms are often reflected also in the spatially-integrated activity of LFPs. As we examined the low frequency components of single-trial MEA raw recordings, a weak signature of the global firing rhythm could be indeed observed (Supplemental Figure S2) although with orders of magnitude worse signal-to-noise ratio and more difficult interpretation [97] than the spike PSTHs.

In conclusion, the emergence of oscillations at physiological gamma frequency and their modulation by the selective activation of excitatory neurons makes large-scale cultured networks, MEAs, and optogenetics relevant for studying and manipulating network phenomena under highly-controlled conditions (see also [85,86,87]), as well as for validating theoretical models of population dynamics with simplified hypotheses on unstructured connection topology.

**Materials and Methods**

**Viral vectors production.** To replace the cytomegalovirus (CMV) promoter in the Adeno-Associated Viral vector (AAV) transfer plasmid [77], a 364 bp fragment of the mouse α-subunit of $Ca^{2+}$/Calmodulin-dependent protein kinase II promoter (CaMKIIα) was PCR-amplified from plasmid 20944 (Addgene, UK), using Fw–AAAGCTAGCACTTGTGGACTAAGTTTGTTCACATCCC– and Rev–AAAAGCGCTGATATCGCTGCCCCCAGAACTAGGGGCCACTCG– as primers. This PCR fragment was ligated in the AAV transfer plasmid, using NheI and Eco47 III, after the CMV promoter was cut out. eGFP was then replaced by CHR2 L132C/T159C-mCherry. The opsin coding sequence was cloned by PCR amplification and the final transfer plasmid was sequence-verified prior to viral vectors production. AAV2/7 viral vectors were produced at the Leuven Viral Vector Core, as described earlier [77]. Briefly, HEK 293T cells (ATCC, Manassas, VA, USA) were seeded in the Hyperflask at $1.2 \times 10^{8}$ cells per viral vector production, in Dulbecco's Modified Eagle's Medium (DMEM) with 2% Fetal Calf Serum. The following day the medium was replaced by serum-free Optimem and cells were transfected with the pAAV transfer plasmid, the rep/cap plasmid, and the pAd.DELTA.F6 plasmid in a 1:1:1 ratio. The supernatant was harvested five days after transient transfection, concentrated using tangential flow filtration, and purified using an iodixanol step gradient. Aliquots were stored at -80°C and quantified, using real-time PCR, as Genome Copies per ml (GC/ml): titers varied between $9.8 \times 10^{11}$ GC/ml and $1.6 \times 10^{12}$ GC/ml.

**Neuronal culturing and transduction.** Primary cultures of mammalian neurons, dissociated from the postnatal rat neocortex, were prepared as in [30], in accordance with international and institutional guidelines on animal welfare. All experimental protocols were specifically approved by the Ethical Committee of the Department of Biomedical Sciences of the University of Antwerpen (permission n. 2011_87), and licensed by the Belgian Animal, Plant and Food Directorate-General of the Federal Department of Public Health, Safety of the Food Chain and the Environment (license n. LA1100469). Briefly, cerebral cortices (excluding the hippocampus) were removed from the brains of Wistar pups (P0), flushed with ice-cold phosphate buffered saline (PBS) containing 20 mM glucose, and roughly minced with a blade. Hemi-cortices were enzymatically digested by adding 0.05 mg/ml trypsin to the solution and by gentle agitation, for 15 min at 37 °C (GFL 1083, Gesellschaft für Labortechnik mbH, Burgwedel, Germany). After stopping the digestion by adding 2 ml of heat-inactivated Normal Horse Serum (NHS), the tissue fragments were left to sediment at room temperature and then mechanically triturated by a 10 ml Falcon pipette, filtered through a nylon monofilament cell strainer (40μm, #352340, BD Falcon, Franklin Lakes, NJ, USA), and centrifuged at room-temperature (220 *g* for 5 min; GS-6, Beckmann-Coulter, Brea, CA, USA). The resulting cell pellet was resuspended in culture medium, composed of 4ml Minimum Essential Medium (MEM), 5% NHS, 50 μg/ml gentamycin, 0.1 mM L-glutamine, and 20 mM sucrose, and diluted to reach a final surface plating density of 6,500 cell/mm$^2$ on glass coverslips (15mm, VD1-0015-Y2MA, Laborimpex, Forest/Vorst, Belgium) and commercial substrate-integrated arrays of microelectrodes (MEAs; Multi Channel Systems GmbH, Reutlingen, Germany).

MEAs with a regular 8x8 arrangement of 60 Titanium Nitrate (TiN) microelectrodes, each with 30μm diameter and 200μm spacing (60MEA200/30iR-ITO-gr, Multi Channel Systems), were employed. Prior to cell seeding, MEAs and coverslips were coated overnight with polyethyleneimine (0.1% wt/vol in millQ water at room temperature), and then extensively washed with milliQ water and let air-drying. Seeded MEAs and coverslips were maintained in an incubator for up to 6 weeks at 37 °C in 5% $CO_2$ and 100% R.H. (5215, Shellab, Cornelius, OR, USA), with MEAs sealed by fluorinated Teflon membranes (Ala-MEA-Mem, Ala Science, Farmingdale, NY, USA) and coverslips stored in Petri dishes.

The culture transduction was performed *in vitro* five days (DIV5) after seeding, by partly replacing the medium of each MEA or coverslip with a 1:50 dilution of viral particles AAV-CaMKIIα-CHR2(LC-TC)-mCherry in fresh medium. At DIV8, fresh and pre-warmed medium was added to reach 1 ml final volume in MEAs and coverslips. From DIV10 on, half of their culture medium volume was replaced every 2 days with fresh, pre-warmed medium. All reagents were obtained from Sigma-Aldrich (St. Louis, MO, USA) or Life Technologies (Gent, Belgium).

**Immunocytochemistry.** Cell identity, density, and opsin expression selectivity and efficiency were evaluated after DIV28 by immunocytochemistry, by imaging mCherry-positive cells and neuronal nuclei (NeuN) stained by an Anti-NeuN monoclonal antibody (clone A60, #MAB377, Millipore, Billerica, MA, USA) (Fig. 1D). Control and transduced cultures on coverslips were washed three times with PBS and fixed by 4% paraformaldehyde dissolved in PBS (10 min at 37 °C). After washing twice with ice-cold PBS, cells membranes were permeabilised (0.25% Triton X-100 in PBS-Tween for 10 min) and pre-incubated with 10% goat serum in PBS-Tween for 30min. Primary Anti-NeuN (1:150) incubation was performed overnight at 4 °C. After repeated PBS washing, the incubation with 1:150 Alexa-488 goat anti-mouse (A11001, Molecular Probes, Eugene, OR, USA) was performed for 1h. After further PBS washing, the coverslips were mounted on glass slides by a medium containing 4′,6-diamidino-2-phenylindole (DAPI), which stained all cell nuclei (Vectashield H-1500, Vector Laboratories, Burlingame, CA, USA).

Several microphotographs of NeuN immunolabelling, DAPI staining, and mCherry red-fluorescence were acquired under epifluorescent microscopy (Olympus DP71, Tokyo, Japan) and processed for semiautomatic cell-counting (ImageJ, NIH, Bethesda, MD, USA), examining four non-overlapping random fields (320μm × 240μm) over at least three distinct coverslips, seeded during two different cell dissociation sessions.

**MEAs extracellular electrophysiological recordings.** Before each experiment, MEAs were transferred from the incubator to an electronic amplifier (MEA-1060-Up-BC, Multi Channel Systems) with 1–3000 Hz bandwidth and amplification factor of 1200, placed inside an electronic-friendly incubator (37 °C, 5% $CO_2$, ~20 R.H.). MEAs were left in the incubator to accommodate for 5min before starting the recordings, and were used to detect non-invasively the spontaneous and (light) evoked neuronal electrical activity. An MCCard A/D board and the MCRack software (Multi Channel Systems) were used to sample (25 kHz/channel), digitize (16 bits), and store on disk the raw voltage traces individually detected by each of the 60 microelectrodes for subsequent analysis. Recordings took place after 28 DIVs, as continuous as well as triggered acquisitions of raw traces. Assessing the health of each MEA, five minutes of spontaneous activity was routinely recorded, before delivering the photo-stimulations (Fig. 1B-C). Light-evoked activity was then recorded across the entire MEA, by monitoring 200ms proceeding and 1000ms following the photo-stimulus onset.

**Intracellular electrophysiological recordings.** Glass coverslips were employed for patch-clamp experiments in sister cultures of at least 20DIV, using an Axon Multiclamp 700B amplifier (Molecular Devices, USA) controlled by the LCG software [78]. Patch pipettes were pulled on a horizontal puller (P97, Sutter, Novato, USA) from filamented borosilicate glass capillaries (World Precision Instrument Inc. (WPI), USA), and had a resistance of 4–5 MΩ, when filled with an intracellular solution containing (in mM): 115 K-gluconate, 20 KCl, 10 HEPES, 4 Mg-ATP, 0.3 $Na_2$-GTP, 10 $Na_2$ –phosphocreatine (pH adjusted to 7.4 with NaOH). An extracellular solution, containing (in mM): NaCl 145, KCl 4, $MgCl_2$ 1, $CaCl_2$ 2, HEPES 5, Na-pyruvate 2, glucose 5 (pH adjusted to 7.4 with NaOH), was constantly perfused at a rate of 1 ml/min and at a temperature of 37±1 °C for each experiment.

*Photo-activated network responses in vitro*Miniature post-synaptic currents (mPSCs) were recorded under voltage-clamp at a hyperpolarized holding potential of -70mV, under the whole-cell configuration from the soma of mCherry-negative neurons, after low-pass filtering at 6 kHz and sampling at 30 kHz the acquired traces. Photo-activated responses were also recorded monitoring the membrane currents, while patching mCherry-positive neurons. Repeated stimulation trials, acquiring 2 s preceding and 2 s following each photo-stimulation were performed, as in MEA experiments. Cells with resting potential above −55 mV, or with a series resistance larger than 25 MΩ, were not included for analysis and discarded. In all the experiments, the Active Electrode Compensation [41] was employed by default through LCG.

**Pharmacology.** In some MEA experiments, Gabazine (GBZ, SR95531) was bath-applied prior to the electrophysiological recordings at the final concentrations of 30 μM, as a selective competitive blocker of $GABA_A$ receptors (Fig. 5). In all intracellular experiments (Fig. 7A-B), 1 μM tetrodotoxin (TTX) was added to the extracellular solution to selectively block sodium channels and thus suppress the generation of action potentials in all neurons. All drugs were obtained from Abcam (Cambridge, England, UK).

**Data Analysis and Statistics.** For MEA recordings, extracellular voltage waveforms of light-evoked neuronal responses were analyzed on-the-fly by custom scripts written in MATLAB (The MathWorks, Natik, MA, USA), while the package QSpike Tools [80] was employed for analyzing spontaneous activity and extracting standard observables such as spike times, spike rate, burst rate, and burst duration [22,43]. Raw voltage traces recorded at each microelectrode (Fig. 1A) were zero-phase band-pass filtered (400-3000Hz) and then processed by an adaptive peak-detection algorithm, based on voltage-threshold crossing and including an elementary sorting of positive and negative amplitude peaks of the corresponding spike waveforms [44-46]. The times of occurrence of extracellular spikes, for each recording channel and each repetition, were stored on disk for subsequent analysis together with their corresponding spoke waveform. Spike waveforms with negative peak voltage amplitudes were sorted and included in the analysis. Specifically, their time stamps were used to estimate network-wide evoked instantaneous spike response probability, computed as a Peri-Stimulus spike Time Histogram (PSTH) over 5 ms bins (Figs. 2A) and whose features were subsequently extracted (Fig. 2B-D).

Spontaneous network bursts were also studied, analyzing the corresponding firing probability over time by computing for each burst a spike time histogram (STH) over 5 ms bins. The overall firing rate profile during a burst, averaged over all the bursts recorded from a MEA, was however not obtained arbitrarily aligning the STHs recorded to the time of their peak [43]. Instead, STHs were first arbitrarily shifted in time and then aligned for averaging, by a series of time intervals that maximized their mutual similarity, through maximizing their correlation coefficients [68].

Frequency-domain analysis was also performed, estimating the spectrogram [47] of each PSTH (estimated over 1 ms bins, Fig. 3A) corresponding to a different photo-stimulus duration (Fig. 3B). In details, the Fast Fourier Transform with a sliding window (50 ms width, 50% overlap) was employed to extract the time-varying spectrum of frequencies contained in the PSTH [48], averaging over the 60 trials and normalizing across frequencies by the PSTH spectrum of the 200 ms preceding the photo-stimulus onset. In addition, the power spectrum computed 75 ms after the stimulus onset (Fig. 3B-C) was conventionally employed to determine existence and location of the *dominant* oscillation frequency, defined arbitrarily as the frequency corresponding to the highest peak of the spectrum, which necessarily exceeded the median value of the spectrum over all frequencies by three times (Fig. 3C).

In intracellular recordings, current traces were low-pass filtered at 1.5 kHz and analysed using custom-written MATLAB scripts to count the number or the instantaneous rate of mPSCs.

Data is presented as mean ± standard error of the mean (SEM), except for Figures 1C and 5A, where box plots show the median (horizontal bar) and the mean values (circle), the 25th and 75th percentiles (box's edges), and the most extreme data points (whiskers). Distributions were compared using the Kolmogorov-Smirnov test and only differences with at least p<0.05 were considered significant. Correlations were assessed by computing the non-parametric Kendall's rank correlation coefficient and its significance level [86].

**Wide-field photo-stimulation.** A blue light-emitting diode (LED) (Rebel, Quadica Development, Canada), powered by an externally dimmable DC driver (LuxDrive, Randolph, USA), was employed to deliver wide-field photo-stimulations. The LED was further equipped with a parabolic reflector lens, to

*Photo-activated network responses in vitro*

collimate the light beam and distribute the emitted power on the bottom of MEAs and coverslips uniformly (Fig. 1A).

In MEA experiments, photo-stimuli were delivered at full power (i.e. 2 mW/mm$^2$, measured at the sample by a calibrated photodiode; 818-ST2-UV, Newport Spectra-Physics, Netherlands) by a voltage-controlled stimulus generator (STG1002, Multi Channel Systems) connected to the DC driver. The generator was programmed over USB to define timing and waveform of each stimulus. 60 identical square pulses were consecutively delivered at a 0.2 Hz repetition rate, and the specific effect of the pulse duration [0.1, 0.5, 1, 2, 5, 10, 20] ms was investigated. Thus, stimulation sessions differed only in the pulse durations, whose value was chosen in a shuffle order across successive sessions.

In the intracellular experiments, an analog output of the acquisition D/A board was connected to the LED DC driver and the LCG software employed to deliver the photo-stimuli as in MEA experiments, repeated 20 times per condition at a 0.2 Hz.

In our experiments, the pulse duration $T$ was varied while keeping its intensity $L_{max}$ to full power to improve proportionality between stimulus and net inward charge, as tested by voltage-clamp experiments (Supplemental Figure S1). In fact, the equation describing the fraction $x$ of open CHR2 channels, from a minimal 2-state (non-inactivating) kinetic biophysical model,

$$1 - x \underset{b}{\overset{a \cdot L}{\rightleftarrows}} x \qquad \dot{x}(t) = -b \cdot x(t) + a \cdot L(t) \cdot [1 - x(t)] \tag{2}$$

predicts a net charge proportional to T, only for a large intensities $L_{max}$.

$$\frac{a \cdot L_{max}}{a \cdot L_{max} + b} \cdot \left[ T + \left( 1 - e^{T \cdot (a \cdot L_{max} + b)} \right) \cdot \frac{a \cdot L_{max} + 2 \cdot b}{b \cdot (a \cdot L_{max} + b)} \right] \approx T \tag{3}$$

**Firing-rate model.** A Wilson and Cowan rate-based mathematical model was defined and numerically simulated [50,51], to review the consequences of recurrent and feed-back interactions between an excitatory (e) and an inhibitory (i) neuronal population. Each population was assumed to be composed of identical and indistinguishable neurons and thus collectively described by a single state variable (i.e. $h_e(t)$ and $h_i(t)$). This variable approximated the total average incoming synaptic inputs to a generic unit of each population. The time-varying output mean firing rate of a population (i.e. E and I) was then assumed, for simplicity, to depend only on its total synaptic input, as in a threshold-linear frequency-current curve (eq. 4):

$$E(t) = \alpha_e [h_e(t) - \theta_e]_+ \qquad I(t) = \alpha_i [h_i(t) - \theta_i]_+, \tag{4}$$

where $[\ ]_+$ indicates the positive part of its argument, θ a minimal activation threshold, and α the slope of the frequency-current response function. In the case of a network topology including recurrent excitation and reciprocal excitation and inhibition (Fig. 6A), the following coupled (non-linear) differential equations fully describe the dynamics of the system:

$$\tau_e \dot{h}_e(t) = -h_e(t) + J_{ee} E(t) - J_{ei} I(t) + E_0 \tag{5}$$
$$\tau_i \dot{h}_{ei}(t) = -h_i(t) + J_{ie} E(t) - J_{ii} I(t), \tag{6}$$

where $J_{ee}, J_{ei}, J_{ie}, J_{ii}$ indicate the average synaptic efficacies of recurrent excitation, reciprocal excitation and inhibition, and mutual inhibition, respectively. For simplicity of the subsequent analysis, we assumed no reciprocal inhibition (i.e. $J_{ii} = 0$, see Figure 6A). $E_0$ represents the time-varying external input current, induced by CHR2 activation, and $\tau_e$ and $\tau_i$ are the time constants of decay of the synaptic currents in the lack of presynaptic activity, for each population respectively. Taking into account the rather slow kinetics of CHR2 LC-TC, $E_0(t)$ was approximated as a brief square pulse followed by a much longer decaying exponential function. By a change of variables, $x = h_e - \theta_e$ and $y = h_i - \theta_i$, and assuming that both $h_e$ and $h_i$ at a given moment are larger $\theta_e$ and $\theta_i$, respectively, eqs. 5-6 can be rewritten as a linear system of differential equations, using linear algebra notation:

$$\begin{bmatrix} \dot{x} \\ \dot{y} \end{bmatrix} = A \begin{bmatrix} x \\ y \end{bmatrix} + b, \qquad \text{where} \tag{7}$$

$$A = \begin{bmatrix} (\alpha_e J_{ee} - 1)/\tau_e & -\alpha_i J_{ei}/\tau_e \\ \alpha_e J_{ie}/\tau_i & -(\alpha_i J_{ii} - 1)/\tau_i \end{bmatrix} \qquad b = \begin{bmatrix} (E_0 - \theta_e)/\tau_e \\ -\theta_i/\tau_i \end{bmatrix} \tag{8}$$

The values of the parameters in A, and in particular of the mean synaptic couplings $J_{ee}, J_{ei}, J_{ie}, J_{ii}$, determine the eigenvalues of the matrix A and thus the dynamics of the output network response (e.g. $x(t)$) to the external stimulus (i.e. $E_0(t)$), being exponential or oscillatory over time [52]. The parameters employed for the simulations of Figure 6 were: $J_{ii} = 0, J_{ei} \in \{0, 12\}$, $J_{ie} = J_{ee} \in \{0.32, 0.61, 0.9, 1.2\}$, $\alpha_i = \alpha_e = 1$, $\theta_e = \theta_i = 0.1$, $\tau_e = \tau_i = 0.1$, $E_0 = 20 \cdot [1 + (u(t-1) - u(t-1.5)) + u(t-1.5) \cdot e^{-(t-1.5)/0.5}]$, where $u(t)$ is equal to 0 for t<0 and equal to 1 otherwise.

*Photo-activated network responses in vitro*

85 Wagenaar, D. A., Madhavan, R., Pine, J. & Potter, S. M. Controlling bursting in cortical cultures with closed-loop multi-electrode stimulation. *The Journal of neuroscience : the official journal of the Society for Neuroscience* **25**, 680-688, doi:10.1523/JNEUROSCI.4209-04.2005 (2005).
86 Fong, M. F., Newman, J. P., Potter, S. M. & Wenner, P. Upward synaptic scaling is dependent on neurotransmission rather than spiking. *Nat Commun* **6**, 6339, doi:10.1038/ncomms7339 (2015).
87 Newman, J. P. *et al.* Optogenetic feedback control of neural activity. *Elife* **4**, doi:10.7554/eLife.07192 (2015).
88 Haroush, N., Marom, S. Slow dynamics in features of synchronized neural network responses. *Frontiers in computational neuroscience* **9**:40, doi:10.3389/fncom.2015.00040 (2015).
89 Droge, M. H., Gross, G. W., Hightower, M. H., & Czisny, L. E. Multielectrode analysis of coordinated, multisite, rhythmic bursting in cultured cns monolayer networks. *The Journal of neuroscience : the official journal of the Society for Neuroscience* **6**, 1583–1592 (1986).
90 Jimbo, Y., Robinson, H. P., & Kawana, A. Simultaneous measurement of intracellular calcium and electrical activity from patterned neural networks in culture. *IEEE Transactions in Biomedical Engineering* **40**, 804–810. doi: 10.1109/10.238465 (1993).
91 Maeda, E., Robinson, H. P., & Kawana, A. The mechanisms of generation and propagation of synchronized bursting in developing networks of cortical neurons. *The Journal of neuroscience : the official journal of the Society for Neuroscience* **15**, 6834–6845 (1995).
92 Shahaf, G., & Marom, S. Learning in networks of cortical neurons. *The Journal of neuroscience : the official journal of the Society for Neuroscience* **21**, 8782–8788 (2001).
93 Wagenaar, D. A., Pine, J., & Potter, S. M. An extremely rich repertoire of bursting patterns during the development of cortical cultures. *BMC Neuroscience* **7**:11. doi: 10.1186/1471-2202-7-11 (2006).
94 Ham, M. I., Bettencourt, L. M., McDaniel, F. D., & Gross, G. W. Spontaneous coordinated activity in cultured networks: analysis of multiple ignition sites, primary circuits, and burst phase delay distributions. *Journal of Computational Neuroscience* **24**, 346–357. doi: 10.1007/s10827-007-0059-1 (2008).
95 Stegenga, J., le Feber, J., & Rutten, W. L. C. Changes within bursts during learning in dissociated neural networks. *Conf. Proc. IEEE Eng. Med. Biol. Soc.* 2008, 4968–4971. doi: 10.1109/IEMBS.2008.4650329 (2008).
96 Einevoll, G. T., Kayser, C., Logothetis, N. K., & Panzeri, S. Modelling and analysis of local field potentials for studying the function of cortical circuits. *Nature Reviews Neuroscience* **14**, 770–785, doi:10.1038/nrn3599 (2013).
97 Waldert, S., Lemon, R. N., & Kraskov, A. Influence of spiking activity on cortical local field potentials. *Journal of Physiology* **591.21**: 5291–5303 (2013).
# Supplementary Information

Authors' preprint 21/23

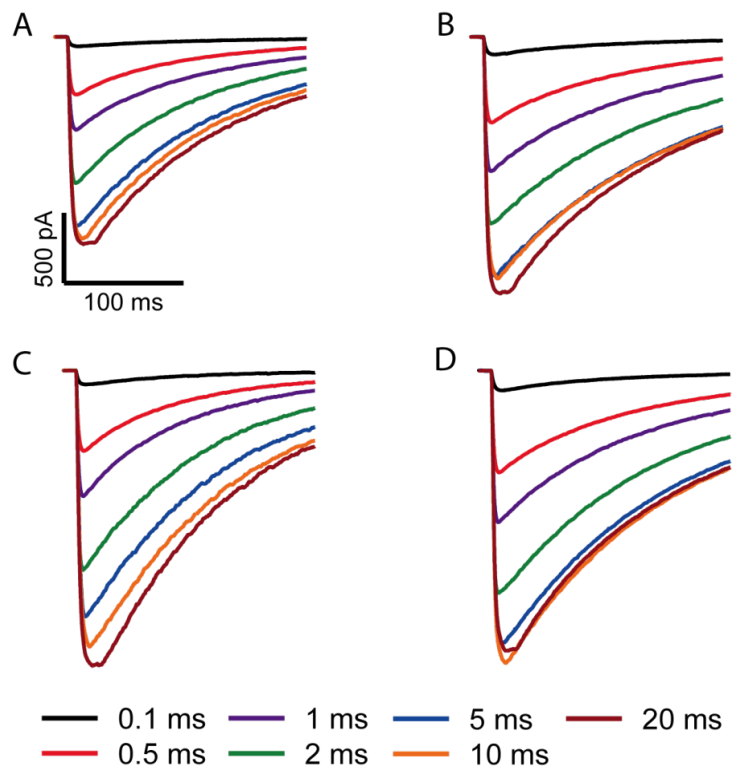

**Figure S1. ChR2-LCTC current dynamics.** Voltage-clamp experiments were performed under TTX in neurons expressing ChR2-LCTC, in order to explore the inward current dynamics. Varying the stimulus duration modulates the peak current amplitude in four representative neurons (**A-D**).

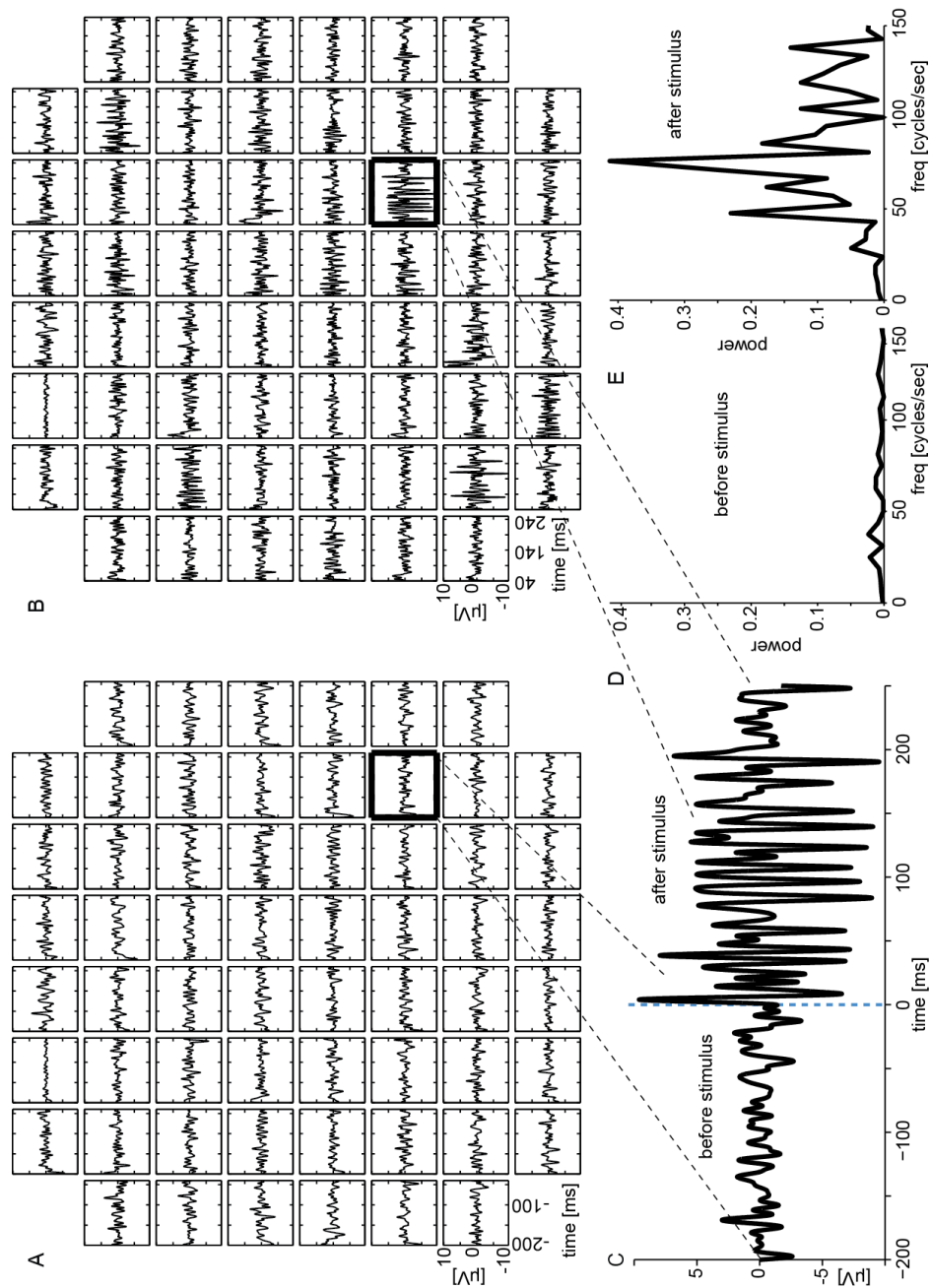

**Figure S2. Light-evoked oscillatory activity is also reflected in single-trial raw voltage traces.** Reminiscent of network rhythm signatures in local field potentials *in vivo* or in brain slices, the impact of brief (1msec) photoactivation was apparent also in single trials raw extracellular voltage recordings, low-pass filtered below 110 cycles/sec. For each MEA microelectrode, filtered voltage traces recorded ~200ms preceding (A) or following (B) the stimulus onset are displayed: the apparent transient increase in signal variance, recorded at most microelectrodes, correspond to reverberating network-wide spiking activity (Figs. 2A, 3). Estimating the power spectrum of a sample low-pass filtered raw trace (C-D) reveals light-induced oscillations in the same range as the spike PSTH (Fig. 3C), although with a much worst signal-to-noise ratio (Fig. 3C).